\documentclass[structabstract]{aa}
\usepackage{graphicx}
\usepackage{txfonts}
\usepackage[breaklinks=true]{hyperref}
\usepackage{natbib}

\usepackage{amsmath}

\newcommand\ensuretext[1]{\ensuremath{\text{#1}}}%
\newcommand\ergcms{\ensuretext{erg\,cm$^{-2}$\,s$^{-1}$}}%
\newcommand\ergs{\ensuretext{erg\,s$^{-1}$}}%
\newcommand\g{\ensuremath{\gamma}}%
\newcommand\fluxone{\ensuremath{(8.60 \pm 2.27) \times 10^{-12}}\,\ergcms}%
\newcommand\fluxtwo{\ensuremath{(1.58 \pm 0.32) \times 10^{-11}}\,\ergcms}%

\begin{document}
   \title{Seyfert~2 galaxies in the GeV band: jets and starburst}

   \author{J.-P. Lenain
          \and
          C. Ricci
          \and
          M. T\"urler
          \and
          D. Dorner
          \and
          R. Walter
          }
   
   \offprints{J.-P.~Lenain\\
     \email{\href{mailto:jean-philippe.lenain@unige.ch}{jean-philippe.lenain@unige.ch}}}
   
   \institute{ISDC Data Centre for Astrophysics, Observatoire de Gen{\`e}ve, Universit{\'e} de Gen{\`e}ve, Chemin d'Ecogia 16, CH-1290 Versoix, Switzerland
   }

   \date{Received 26 August 2010 / Accepted 23 September 2010}

 
  \abstract
   {The \textit{Fermi}/LAT collaboration recently reported the detection of starburt galaxies in the high energy \g-ray domain, as well as radio-loud narrow-line Seyfert\,1 objects.}
   {Motivated by the presence of sources close to the location of composite starburst/Seyfert~2 galaxies in the first year \textit{Fermi}/LAT catalogue, we aim at studying high energy \g-ray emission from such objects, and at disentangling the emission of starburst and Seyfert activity.}
   {We analysed 1.6 years of {\it Fermi}/LAT data from NGC\,1068 and NGC\,4945, which count among the brightest Seyfert~2 galaxies. We search for potential variability of the high energy signal, and derive a spectrum of these sources. We also analyse public \textit{INTEGRAL} IBIS/ISGRI data over the last seven years to derive their hard X-ray spectrum.}
   {We find an excess of high energy \g-rays of 8.3\,$\sigma$ and 9.2\,$\sigma$ for 1FGL\,J0242.7$+$0007 and 1FGL\,J1305.4$-$4928, which are found to be consistent with the position of the Seyfert 2 galaxies NGC\,1068 and NGC\,4945, respectively. The energy spectrum of the sources can be described by a power law with a photon index of $\Gamma=2.31 \pm 0.13$ and a flux of $F_\mathrm{100\,MeV-100\,GeV}=$\fluxone\ for NGC\,1068, while for NGC\,4945, we obtain a photon index of $\Gamma=2.31 \pm 0.10$ and a flux of $F_\mathrm{100\,MeV-100\,GeV}=$\fluxtwo. For both sources, we detect no significant variability nor any indication of a curvature of the spectrum. While the high energy emission of NGC\,4945 is consistent with starburst activity, that of NGC\,1068 is an order of magnitude above expectations, suggesting dominant emission from the active nucleus. We show that a leptonic scenario can account for the multi-wavelength spectral energy distribution of NGC\,1068.
}
   {High energy \g-ray emission is revealed for the first time in a Seyfert~2 galaxy. If this result is confirmed in other objects, new perspectives would be opened up into the GeV band, with the discovery of a new class of high energy \g-ray emitters.}

   \keywords{gamma rays: galaxies --
                galaxies: Seyfert --
                galaxies: individual: (NGC\,1068, NGC\,4945) --
                radiation mechanisms: non-thermal
               }

   \maketitle
%

\section{Introduction}

Somewhat unexpectedly, the {\it Fermi}/LAT collaboration reported the discovery of four radio-loud narrow-line Seyfert 1 galaxies \citep{2009ApJ...699..976A,2009ApJ...707L.142A}, suggesting that these objects, previously undetected at high energies, could constitute a new class of high energy emitters. Extending this idea to Seyfert 2 galaxies, we show here that two such active galactic nuclei (AGN) among the closest and brightest in the X-ray sky, NGC\,1068 and NGC\,4945, are detected at high energies using {\it Fermi}/LAT data. We investigate whether this emission is dominated by AGN or starburst activity.

\object{NGC\,1068} is an archetypal Seyfert 2 galaxy, located at $z=0.003786$, i.e. 14.4\,Mpc away, and harbours a hidden Seyfert 1 core. It was used by \citet{1985ApJ...297..621A} to propose the AGN unification. Given the proximity of this spiral galaxy, its extension is well observable in visible light, and it is the closest, as well as one of the brightest, Seyfert 2 galaxy. This source exhibits both AGN and starburst activities in its central region \citep[e.g.][]{1987ApJ...321..755L,2004Natur.429...47J}. A dusty torus lies in its central part, dominating the soft X-ray emission by reflection from the central nucleus. A circumnuclear starburst region is located at $\sim$1\,kpc from the core, dominating the infrared emission of the broadband spectral energy distribution (SED) \citep{1989ApJ...343..158T}.

\citet{2006ApJ...638..120C} studied the morphology of the inner warped disc, likely due to the Bardeen-Petterson effect, using VLBA observations. \citet{2008A+A...477..517C} showed that the emission at 43\,GHz from the so-called S1 central radio component is more likely dominated by thermal emission from the hot inner region of the obscuring torus, although \citet{2004ApJ...613..794G} pointed out that the North-East radio component could be dominated by synchrotron emission. This is corroborated by the results of \citet{2008A+A...485...33H}, who argued that the radio emission from the core could be dominated by synchrotron or free-free emission, while near-IR data are dominated by the thermal emission from the dusty torus. Thanks to near-IR interferometric data using the MIDI instrument at the VLT, \citet{2009MNRAS.394.1325R} found the torus to be composed of two thermal components at 300\,K and 800\,K.

In the high energy domain, \textit{Chandra} observations of the core of NGC\,1068 \citep{2003A+A...402..849O} showed that the X-ray emission is due to photoionisation in the extended, clumpy narrow line region. \citet{2004A+A...414..155M} showed that the neutral reflector is Compton-thick, using \textit{XMM}-Newton observations.

%
\begin{figure*}
  \centering
  \includegraphics[width=0.98\columnwidth]{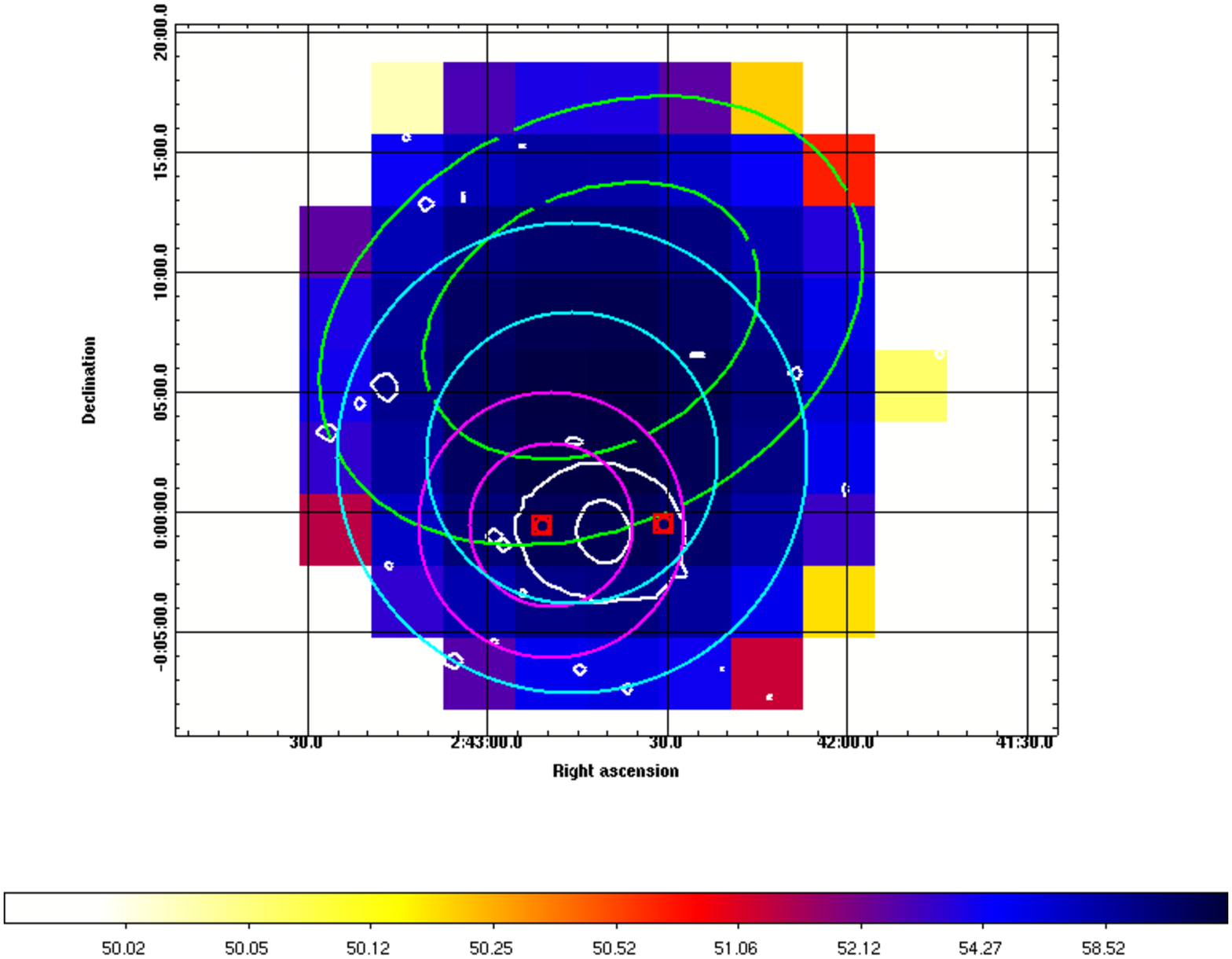}
  \includegraphics[width=0.98\columnwidth]{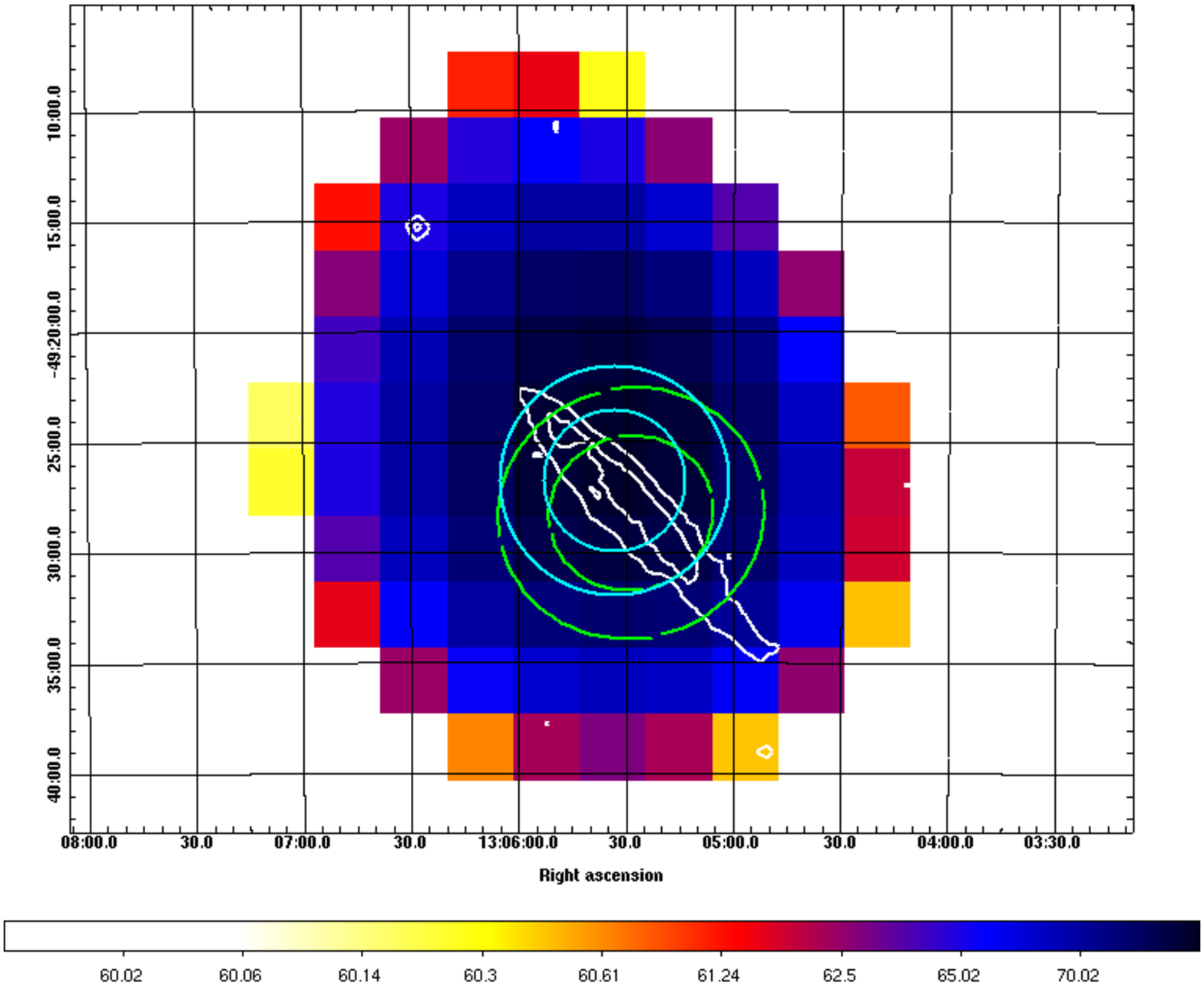}
  \caption{{\it Left}: TS map of NGC\,1068 between 100\,MeV and 100\,GeV. The green ellipses show the 68\% and the 95\% position errors from the 1FGL catalogue, the cyan and magenta circles show respectively the position error (at 68\% and 95\% CL) for the full data set with all the events accounted for, and for front events only. The white contours are taken from an optical image from the Digital Sky Survey, showing the extent of the Seyfert galaxy. The red boxed points denote the position of the two quasars nearby NGC\,1068 (see text in Section~\ref{sec-discussion} for more details). {\it Right}: same as left panel, for NGC\,4945. For clarity, we only present here the position error circle for all events.
  }
  \label{fig-TSmap}
\end{figure*}

Motivated by the presence of a \textit{Fermi}/LAT source, 1FGL\,J0242.7$+$0007, in the region of NGC\,1068 in the 11-months \textit{Fermi}/LAT catalogue \citep[1FGL,][]{2010ApJS..188..405A}, with no proposed counterpart in radio nor in \g-rays, we analyse here 1.6 years of data from the {\it Fermi}/LAT instrument, in order to better constrain the origin and properties of the \g-ray emission in this region.


\object{NGC\,4945} is also a Seyfert 2 galaxy at $z=0.001908$ exhibiting starburst activity in its central region \citep{1993ApJ...409..155I,1994ApJ...429..602M}. Its emission extends up to soft $\gamma$-rays \citep{2009A+A...507..549P}, as observed by \textit{INTEGRAL}/SPI. It is one of the brightest hard X-ray AGN \citep[see][and references therein, and Ricci et al., in prep.]{2008PASJ...60S.251I}. NGC\,4945 was found to be a Compton thick AGN, based on \textit{GINGA} \citep{1993ApJ...409..155I} and {\it INTEGRAL} observations \citep{2009A+A...505..417B}. This source was already reported as a high-energy \g-ray emitter by the \textit{Fermi}/LAT collaboration in the 11-months catalogue \citep{2010ApJS..188..405A}, although the authors did not conclude whether this high energy emission is due to starburst or AGN activity. 

We present here a detailed analysis of 1.6 years of \textit{Fermi}/LAT data of NGC\,4945 to compare the results to those of NGC\,1068.


\section{\textit{Fermi}/LAT data analysis}
\label{sec-fermi_analysis}

We present in the following our analysis of \textit{Fermi}/LAT data on the sources 1FGL\,J0242.7$+$0007 and 1FGL\,J1305.4$-$4928 reported in the 1FGL catalogue. We will show that these sources can be associated to the Seyfert 2 objects NGC\,1068 and NGC\,4945, respectively.

We analysed $\sim$1.6\,yr of {\it Fermi}/LAT data, spanning from August 4, 2008 to March 15, 2010, from a region of interest of 10\degr\ in radius around NGC\,1068, using the publicly available {\it Science Tools}\footnote{\href{http://fermi.gsfc.nasa.gov/ssc/data/analysis/software/}{http://fermi.gsfc.nasa.gov/ssc/data/analysis/software/}}, and we followed the unbinned likelihood analysis scheme presented in \citet{2009ApJ...697.1071A}. We always used the so-called ``diffuse class'' events, which are the events detected by the {\it Fermi}/LAT with the highest probability to be \g-ray photons, and the P6V3 instrument response.

Using the {\it gtlike} tool and assuming a power-law shape for the source spectrum, the Test Statistic \citep[TS,][]{1996ApJ...461..396M} of the likelihood analysis is 68.6, corresponding approximately to a 8.3\,$\sigma$ source detection in the 100\,MeV--100\,GeV range. The corresponding TS map is shown in Fig.~\ref{fig-TSmap}. The best-fit location of the source using the {\it gtfindsrc} tool is $\alpha_{J2000} = 2^\mathrm{h} 42^\mathrm{m} 46^\mathrm{s}$, $\delta_{J2000} = 0\degr 2\arcmin 14\arcsec$ with an error circle radius of $\sim$6\farcm1 (68\% confidence level, CL), and is fully compatible with the position reported in the 1FGL catalogue. The maximum photon energy detected from the source is 20.0\,GeV, located at 0\farcm48 from the position of NGC\,1068.

Given the energy dependence of the point spread function (PSF) of {\it Fermi}/LAT, we performed a second analysis, using only {\it front} events for which the PSF is narrower\footnote{see \href{http://www-glast.slac.stanford.edu/software/IS/glast_lat_performance.htm}{http://www-glast.slac.stanford.edu/software/IS/glast\_lat\_\\performance.htm}}. Front events are those converted in the top layers of the tracker of the LAT instrument \citep[see][for more details]{2009ApJ...697.1071A}. In this latter analysis, the TS of the source is 42, still sufficient to derive a position of the {\it Fermi}/LAT excess. The best-fit position of the source is then $\alpha_{J2000} = 2^\mathrm{h} 42^\mathrm{m} 49^\mathrm{s}$, $\delta_{J2000} = -0\degr 0\arcmin 30\arcsec$, with an error circle radius of $\sim$3\farcm4 (68\% CL), which is only 2\farcm1 away from the nominal position of NGC\,1068. This latter result on the position is only marginally compatible with the position reported in the 1FGL catalogue.

Given the angular distance between the {\it Fermi} source 1FGL\,J0242.7$+$0007 and NGC\,1068, and its optical extension of $\sim$6\farcm5, we propose that this {\it Fermi} source is actually associated with the Seyfert 2 galaxy NGC\,1068.

All the sources reported in the {\it Fermi}/LAT 11-months catalogue \citep{2010ApJS..188..405A} within a radius of 15\degr\ around NGC\,1068 were included in the likelihood analysis, and modelled with power-law spectra. We first identify the sources not contributing significantly to the likelihood and remove them from the model. The spectral parameters of the remaining sources are left freely varying. For NGC\,1068, we obtain $F_\mathrm{100 MeV - 100 GeV} =$\fluxone\ and $\Gamma = 2.31 \pm 0.13$. We also tried to fit the data with a broken power-law or a log parabola, but this did not improve the likelihood.

We performed two other analyses, focussing on the 1--100\,GeV band, for all events and for front events only. This energy band benefits from a better PSF compared to lower energy events, despite a much smaller count rate due to the soft spectrum of the source. We obtain consistent results to those reported above, including comparable position error on the source.

\begin{figure}
  \centering
  \includegraphics[width=\columnwidth]{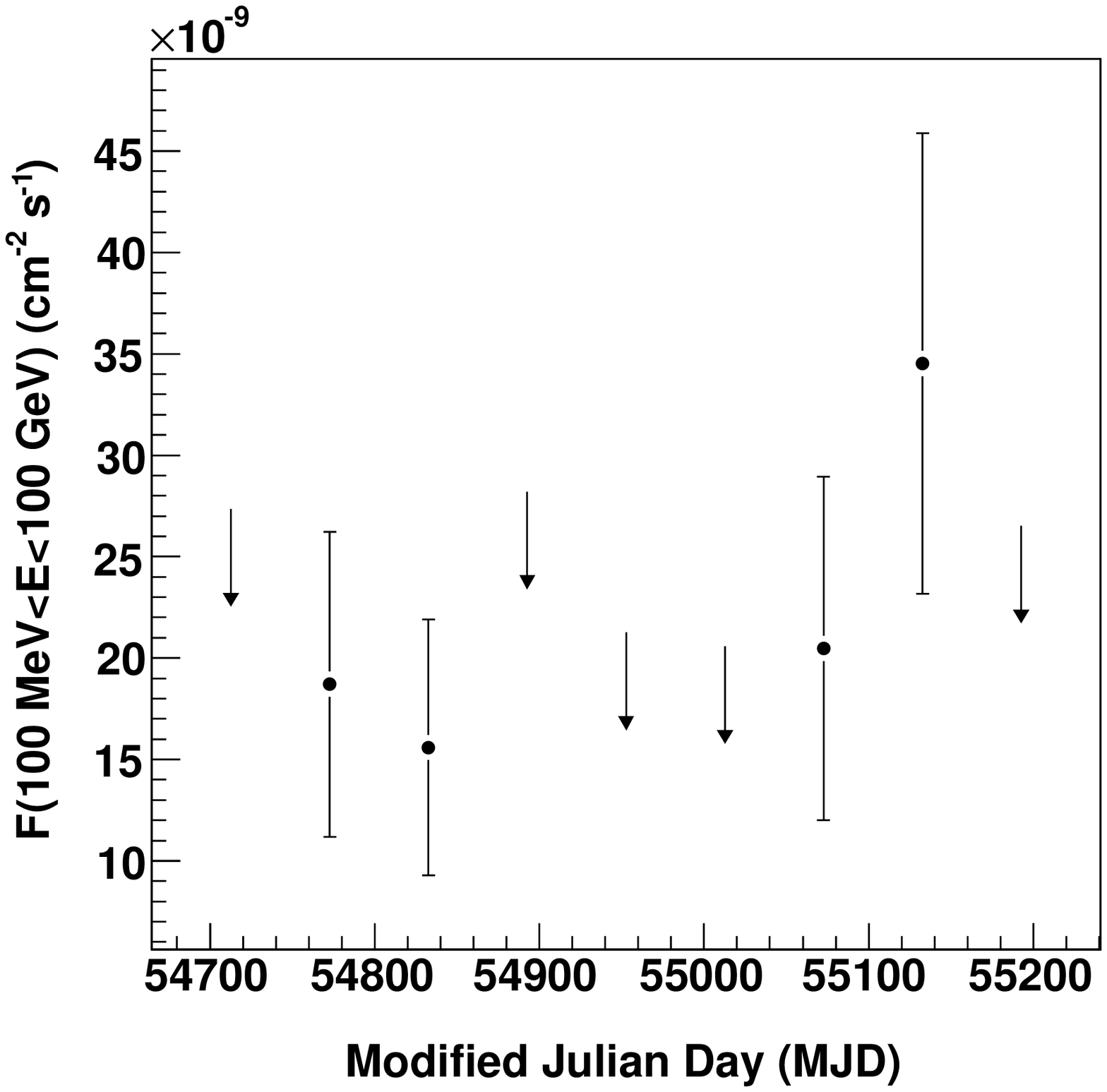}
  \includegraphics[width=\columnwidth]{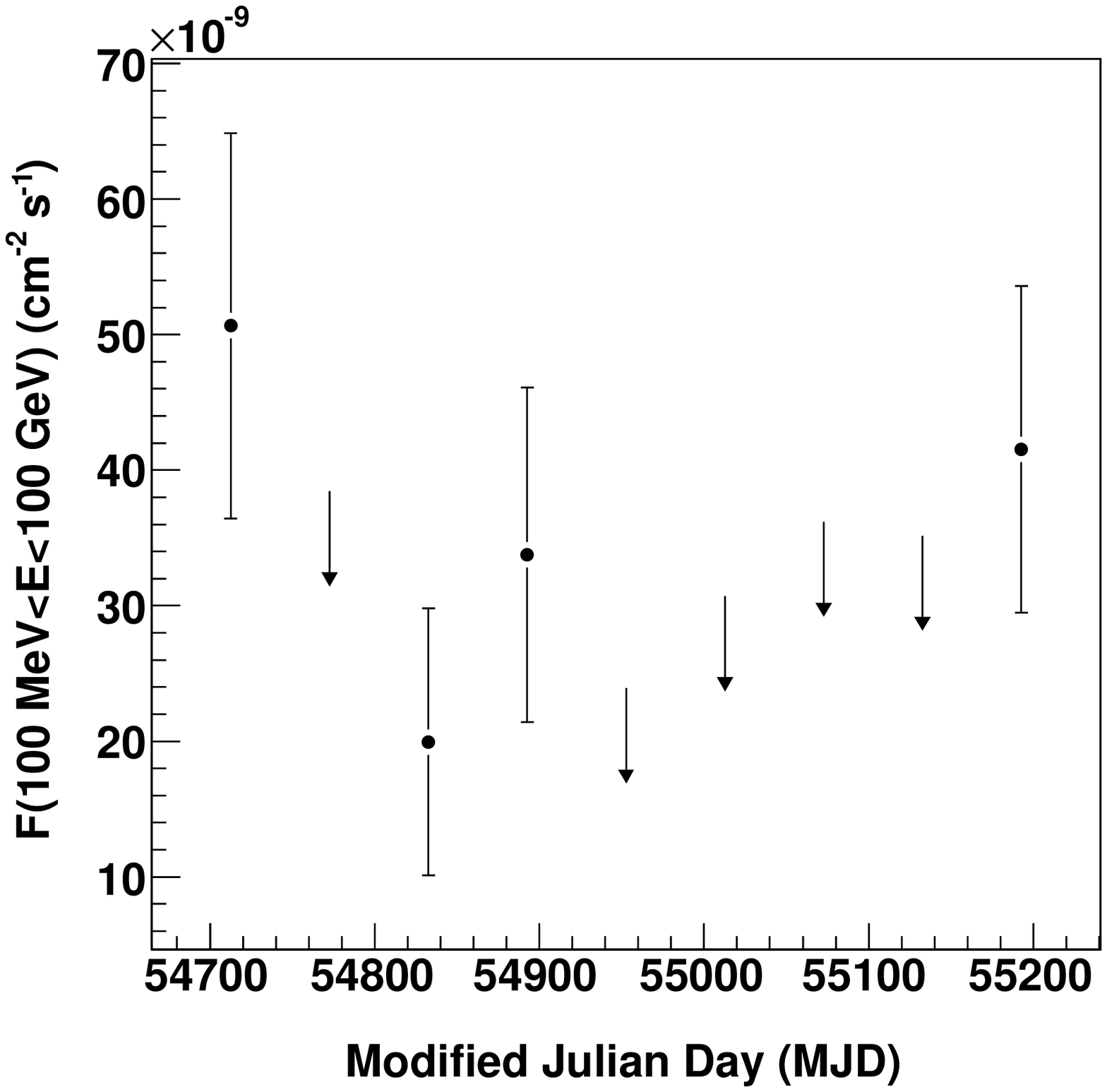}
  \caption{\textit{Top:} Light curve of NGC\,1068 in the 100\,MeV--100\,GeV energy band. The time bins are 60 days wide, and the arrows represent 95\% CL upper limits. \textit{Bottom:} same as top for NGC\,4945. In both sources, no significant variability is visible.
  }
  \label{fig-LC}
\end{figure}

We also investigated the potential variability, performing a likelihood analysis in different time intervals, for a time bin of 60 days, letting only the normalisation of the source free to vary. As shown in Fig.~\ref{fig-LC}, no significant variability was detected.

%


We followed the same procedure for the analysis of {\it Fermi}/LAT data in the region of NGC\,4945, for the source 1FGL\,J1305.4$-$4928. The likelihood analysis on NGC\,4945 results in a TS of 85.3 in the 100\,MeV--100\,GeV band, equivalent to a 9.2\,$\sigma$ detection. Assuming a power-law shape on the source energy spectrum, a photon index of $\Gamma= 2.31 \pm 0.10$ and a flux of $F_\mathrm{100 MeV-100 GeV}=$\fluxtwo\ are found. The highest energy photon detected is 20.7\,GeV, located at 0\farcm22 from NGC\,4945. As for NGC\,1068, the use of a broken power-law or a log parabola did not improve the likelihood, and no significant variability was found in the data (see Fig.~\ref{fig-LC}), which are statistically consistent with a constant for both sources.

The analysis of the whole data set gives a position of $\alpha_{J2000} = 13^\mathrm{h} 05^\mathrm{m} 33^\mathrm{s}$, $\delta_{J2000} = -49\degr 26\arcmin 44\arcsec$ with an error circle radius of 3\farcm2 (68\% CL), only 1\farcm6 away from the nominal position of NGC\,4945, while the analysis for the front events only results in a position of $\alpha_{J2000} = 13^\mathrm{h} 05^\mathrm{m} 34^\mathrm{s}$, $\delta_{J2000} = -49\degr 26\arcmin 49\arcsec$ with an error circle radius of 3\farcm3 (68\% CL). The source position is fully compatible with the results reported in the 1FGL catalogue.

It should also be noted that for both sources, the spectral parameters found are fully compatible with the ones reported in the 1FGL catalogue.

\section{\textit{INTEGRAL} data analysis}
\label{sec-integral_analysis}


For the extraction of the IBIS/ISGRI spectra, we used all the public data obtained by {\it INTEGRAL} as of May 2010, for a total of 564 and 865 pointings (ScWs) on NGC\,1068 and NGC\,4945, respectively. The typical exposure of each pointing is $(1-3)\times 10^3$\,s, and only ScWs with an effective exposure longer than 200\,s were kept, spanning times between December~30, 2002 (revolution 26) and April~7, 2009 (revolution 791). The total exposure is of 622 ks for NGC\,1068, and of 927 ks for NGC\,4945. In the large 17--80 keV band, NGC\,1068 and NGC\,4945 were detected with a significance of 14.0\,$\sigma$ and 138.8\,$\sigma$, respectively.

IBIS/ISGRI has a large field of view of $29\degr \times 29\degr$ with a spatial resolution of 12 arcmin. Note that because of the nature of coded mask imaging the whole sky image taken by the instrument has to be considered in the analysis, because all sources in the field of view contribute to the signal \citep{1987SSRv...45..349C}.

The {\it ISGRI} data were reduced using the {\it INTEGRAL} Offline Scientific Analysis software\footnote{\href{http://www.isdc.unige.ch/integral/}{http://www.isdc.unige.ch/integral/}} version 9.0, publicly released by the {\it INTEGRAL} Science Data Centre \citep{2003A+A...411L..53C}. The analysis of the IBIS/ISGRI data is based on a cross-correlation procedure between the recorded image on the detector plane and a decoding array derived from the mask pattern. We created mosaic images of all pointings in 10 energy bins, and extracted the spectra using {\tt mosaic\_spec}.

We used the latest detector redistribution matrix files (RMF) and calculated the ancillary response functions (ARFs) with a weighted average of the 9 available ARFs, based on the number of ScWs within the validity time of a particular ARF.


We have also checked the \textit{Swift}/BAT \citep{2005SSRv..120..143B} spectra of the two sources. The 18 months \textit{Swift}/BAT spectra have been extracted from the BAT archive \citep{2010A+A...510A..47S} served by the HEAVENS source results archive (Walter et~al., in prep.). The ISGRI and BAT spectra are in good agreement for both sources.

A simple power-law fit to the combined \textit{INTEGRAL}/ISGRI  and \textit{Swift}/BAT data of NGC\,1068 using \texttt{Xspec} results in $\chi^2/$d.o.f.=$10.21/11$. The resulting photon index is $\Gamma=2.08^{+0.21}_{-0.19}$, and the flux density is $F_{17-80\,\mathrm{keV}}=1.86^{+0.15}_{-1.23} \times 10^{-11}$\ergcms (both 90\% CL).

The hard X-ray spectrum for the combined ISGRI and BAT data of NGC\,4945 is not well reproduced by a simple power-law, and a more complex model is required. A power-law with exponential cut-off yields $\chi^2/$d.o.f.=$89.6/9$, which is not satisfactory, even though the data clearly show a curvature. Instead, an absorbed power-law yields $\chi^2/$d.o.f.=$54.77/9$, with a photon index of $\Gamma=2.03 \pm 0.04$, and an intrinsic absorption column $N_H=(6.66 \pm 0.45) \times 10^{24}$\,cm$^{-2}$. Even though the $\chi^2$ is not entirely satisfactory, we will use the latter spectral model to build the broadband SED of NGC\,4945, because it provides a better description of the data. A detailed analysis of the hard X-ray spectral shape is beyond the scope of this paper.

%

\section{Discussion}
\label{sec-discussion}

\begin{figure}
  \centering
  \includegraphics[width=\columnwidth]{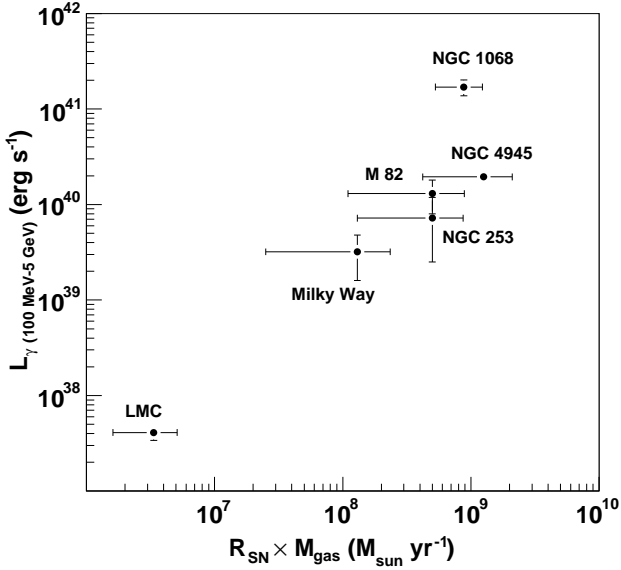}
  \caption{Relationship between SN rate, total gas mass and \g-ray luminosity of NGC\,1068, NGC\,4945, NGC\,253, M\,82, the LMC and the Milky Way.}
  \label{fig-SNrateMgas_GammaLumin}
\end{figure}

It shall be noted that two quasars are present in the field of view of NGC\,1068. The quasar SDSS J024230.65-000029.6 ($z=2.51$) lies $\sim$2\farcm4 from NGC\,1068, and another quasar, SDSS J024250.98-000031.6 ($z=2.18$, a.k.a. QSO\,0240$-$0012), is $\sim$2\farcm6 away from NGC\,1068. However, given their high redshift, it is unlikely that one of these objects could be at the origin of the high energy emission, although this can not be completely excluded. Indeed according to \citet{2010ApJ...715..429A}, the highest redshift source detected by the \textit{Fermi}/LAT is a flat spectrum radio quasar with a redshift of $z=3.10$. Given the similarities of the high energy spectra of 1FGL\,J0242.7$+$0007 and NGC\,4945, it appears however more likely that 1FGL\,J0242.7$+$0007 is associated with NGC\,1068.

\begin{table*}
\caption{Comparison of the starburst properties of NGC\,1068 with NGC\,4945, NGC\,253, M\,82, the LMC and the Milky Way.}
\label{tab-compar}
\centering
\begin{tabular}{lcccccccc}
\hline\hline
Source & $d$   & $\Gamma$\tablefootmark{a} & $L_\gamma$\tablefootmark{a}         & $R_\mathrm{SN}$\tablefootmark{b} & $M_\mathrm{gas}$\tablefootmark{c} & $L_\mathrm{IR}$ & $L_\mathrm{5\,GHz}$ & $L_\gamma$/$L_\mathrm{5\,GHz}$\\
       & (Mpc) &          & ($10^{39}$\,\ergs) & (yr$^{-1}$) & ($10^9 M_\odot$) & ($10^{44}$\,\ergs) & ($10^{38}$\,\ergs)\\
\hline
   NGC\,1068 & 14.4  & $2.31 \pm 0.13$ & $170  \pm 32$     & $0.20 \pm 0.08$   & $4.4$           & 2.7   & 16.6 & 102\\
   NGC\,4945 & 3.6   & $2.30 \pm 0.10$ & $19.5 \pm 2.9$    & 0.1--0.5          & $4.2$           & 1.4   & 2.2 & 89\\
   NGC\,253  & 3.9   & $1.95 \pm 0.4$  & $7.2  \pm 4.7$    & $0.2 \pm 0.1$     & $2.5 \pm 0.6$   & 1.0   & 2.0 & 36\\
   M\,82     & 3.6   & $2.2  \pm 0.2$  & $13.0 \pm 5.0$    & $0.2 \pm 0.1$     & $2.5 \pm 0.7$   & 1.2   & 3.0 & 43\\
   LMC       & 0.049 & --              & $0.041 \pm 0.007$ & $0.005 \pm 0.002$ & $0.67 \pm 0.08$ & 0.016 & 0.017 & 24\\
   Milky Way & --    & --              & $3.2 \pm 1.6$     & $0.02 \pm 0.01$   & $6.5 \pm 2.0$   & 0.5   & 0.36 & 89\\
\hline
\end{tabular}
\tablefoot{\\
\tablefoottext{a}{Photon indices and luminosities in the \g-ray band from \citet{2010ApJ...709L.152A} for M\,82, NGC\,253, the LMC and the Milky Way. The \g-ray luminosities are given in the 100\,MeV--5\,GeV band, and the photon indices $\Gamma$ are given above 200\,MeV for an easier comparison with the results from \citet{2010ApJ...709L.152A}.}\\
\tablefoottext{b}{SN rate estimates from \citet{1982ApJ...263..576W,1994ApJ...421...92B,2003A+A...401..519M} for NGC\,1068, from \citet{2009AJ....137..537L} for NGC\,4945, and from \citet[][and references therein]{2010ApJ...709L.152A} for M\,82, NGC\,253, the LMC and the Milky Way.}\\
\tablefoottext{c}{Gas mass estimates from \citet{1990ApJ...351..422S} for NGC\,1068, from \citet{2008A+A...490...77W} for NGC\,4945, and from \citet[][and references therein]{2010ApJ...709L.152A} for M\,82, NGC\,253, the LMC and the Milky Way.}
}
\end{table*}

\begin{figure}
  \centering
  \includegraphics[angle=-90,width=\columnwidth]{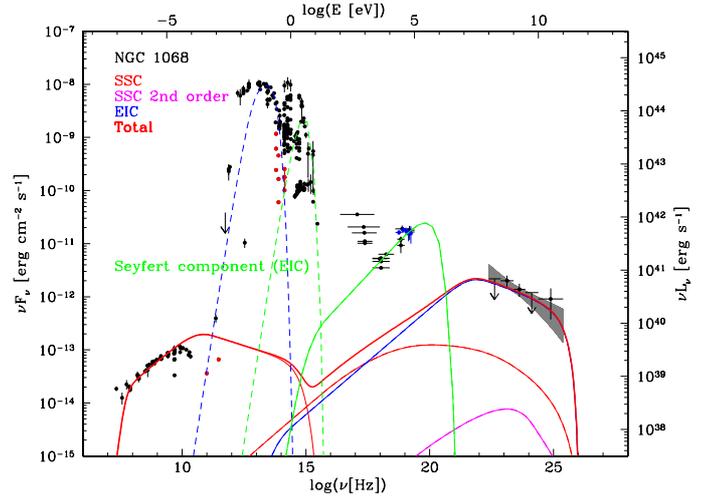}
  \caption{Spectral energy distribution of NGC\,1068, including the {\it Fermi}/LAT spectrum. The black and red points are archival data from the NED, the red ones denote data taken from the central region of NGC\,1068. For clarity, we only show the \textit{INTEGRAL} IBIS/ISGRI data in blue in the hard X-rays. The EIC model for the outflow is shown in blue, and the corresponding SSC emission is shown in thin red and magenta lines for first and second order components, respectively. The thick red line shows the sum of the different emission components from the large outflow. The EIC component from the accretion disc is shown in green.
  }
  \label{fig-model-NGC1068}
\end{figure}

\begin{table*}
\caption{Model parameters for NGC\,1068 and NGC\,4945.}
\label{tab-model_parameters}
\centering
\begin{tabular}{lcccccccccccc}
\hline\hline
Source & Component\tablefootmark{a} & $\delta_\mathrm{b}$ & $B$ (G)  & $r_\mathrm{b}$ (cm) & $T$ (K) & $\tau L_\mathrm{nuc}$ (\ergs) & $R$ (cm) & $K$ (cm$^{-3}$) & $n_1$ & $n_2$ & $\gamma_\mathrm{break}$ & $\gamma_\mathrm{max}$\\
\hline
NGC\,1068 & 1 & 1.2 & $10^{-4}$ & $2.0 \times 10^{19}$ & 130--520 & $1.5 \times 10^{42}$ & $2.2 \times 10^{20}$& 12.5  & 2.2 & 3.3 & $10^4$ & $10^6$\\
NGC\,1068 & 2 & 1.2 & 1        & $10^{15}$            & $10^4$    & $3.0 \times 10^{41}$ & $5 \times 10^{14}$ & $10^6$ & 2.0 & --  & --     & 300\\
NGC\,4945 & 2 & 1.2 & 1        & $2.6 \times 10^{15}$ & $10^4$   & $5.0 \times 10^{41}$ & $5 \times 10^{15}$ & $10^6$ & 1.5 & --  & --      & 160\\
\hline
\end{tabular}
\tablefoot{\\
\tablefoottext{a}{The component 1 refers to the model for the large outflow, while component 2 points to the model for the Seyfert emission in hard X-rays from the accretion disc.}
}
\end{table*}

%
\begin{figure}
  \centering
  \includegraphics[angle=-90,width=\columnwidth]{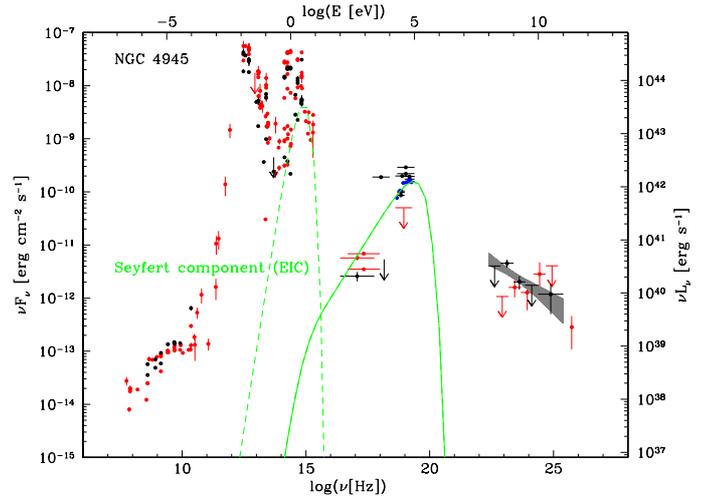}
  \caption{Spectral energy distribution of NGC\,4945, including the {\it Fermi}/LAT spectrum (black points). For clarity, we only show the \textit{INTEGRAL} IBIS/ISGRI data in blue in the hard X-rays. The model for the EIC component from the accretion disc is shown in green. We show in red the data of NGC\,253 as taken from the NED, with the \textit{Fermi}/LAT spectrum published in \citet{2010ApJ...709L.152A} as well as the H.E.S.S. flux measurement from \citet{2009Sci...326.1080A}, for comparison. The two objects have clearly very similar SEDs. The luminosity axis on the right is given for NGC\,4945.
  }
  \label{fig-SED-NGC4945}
\end{figure}


An obvious test to rule out the starburst origin of the high-energy emission would be to detect significant \g-ray variations from one of these objects, as the emission arising from starburst activity is expected to be steady. Due to the lack of statistics, no conclusions can be drawn about the variability of NGC\,1068 or NGC\,4945 from their light curves (see Fig.~\ref{fig-LC}).

Another way to disentangle the starburst or AGN origin of the high-energy emission from these galaxies is to compare their \g-ray luminosity with those of the famous starburst galaxies \object{NGC\,253} and \object{M\,82}, which are also detected in the very high energy domain by H.E.S.S. \citep{2009Sci...326.1080A} and VERITAS \citep{2009Natur.462..770V}, respectively. Both have a luminosity in the 100\,MeV--5\,GeV band of the order of $\approx$10$^{40}$\,\ergs, as detected with {\it Fermi} \citep[][]{2010ApJ...709L.152A}. Computing the luminosities of NGC\,1068 and NGC\,4945 in the same energy band, for comparison, we obtain $1.7 \times 10^{41}$\,\ergs\ and $2.0 \times 10^{40}$\,\ergs, respectively.

Following \citet{2010ApJ...709L.152A}, we compare the supernova rate $R_\mathrm{SN}$, the total gas mass $M_\mathrm{gas}$ and the \g-ray luminosities of NGC\,1068 and NGC\,4945 to the ones of NGC\,253, M\,82, the Large Magellanic Cloud (\object{LMC}) and the \object{Milky Way}, as well as their infrared and radio luminosities (see Table~\ref{tab-compar}). These objects are the only extragalactic sources  which are not AGN known to emit high energy \g-rays. Models attributing the \g-rays to cosmic-ray processes, as expected in a starburst galaxy, depend on the product $M_\mathrm{gas} R_\mathrm{SN}$. Adding NGC\,1068 and NGC\,4945 to this picture \citep[see Fig.~2 in][]{2010ApJ...709L.152A}, and assuming that their \g-ray emission is entirely due to starburst activity, confirms the general trend, but we only find a correlation coefficient of 0.44 for a linear relationship between $M_\mathrm{gas} R_\mathrm{SN}$ and $L_\g$ (see Fig.~\ref{fig-SNrateMgas_GammaLumin}), corresponding to a null hypothesis probability as high as 38\%, although the number of considered sources is obviously too small to definitely conclude.

The \g-ray luminosity and the SN rate of NGC\,4945 are fully consistent with those of NGC\,253 and M\,82, hence even though this object is a composite starburst/AGN, its high energy emission detected using {\it Fermi} could be explained only in terms of its starburst activity. Concerning NGC\,1068, the situation is more complex. Its SN rate is comparable to those of M\,82 and NGC\,253, but its radio and \g-ray luminosities are higher by a factor $\sim$10. This would suggest that its high energy \g-ray emission is more likely dominated by the central AGN activity. Indeed, removing NGC\,1068 from the sample of sources, the correlation coefficient for the relationship between $M_\mathrm{gas} R_\mathrm{SN}$ and the \g-ray luminosity becomes 0.95, with a null hypothesis probability of only $\sim$1\%. This also tends to prove the peculiar role of NGC\,1068 in this study. This is also strengthened by the fact that radio maps of NGC\,1068 clearly show a structured jet, on parsec- and kiloparsec-scales, modelled by the outflow from the central AGN \citep[see e.g.][]{2004ApJ...613..794G,2006AJ....132..546G}. On the contrary, the radio morphology of NGC\,4945 shows an extended emission consistent with the optical morphology tracing the edge-on galaxy \citep[see e.g.][]{1992ApJS...80..137J}, indicating a starburst emission.


Assuming that the high energy emission of NGC\,1068 is indeed due to the AGN activity, and taking a scenario similar to what was proposed in \citet{2008A+A...478..111L} where the jet is misaligned with respect to the line of sight, we consider here the possibility that the outflow in NGC\,1068 could be a high energy emitter.

A large, mildly-relativistic zone of the wind-like outflow, at a few tenth of parsecs from the core, could emit high-energy \g-rays through external inverse Compton process (EIC) \citep[see e.g.][]{1987ApJ...322..650B}. At such distances, the infrared photon energy density is still high enough to ensure a significant emission while being not too important to prevent high optical opacity from pair production. At about 100 parsecs from the core, the magnetic field strength is also expected to be low, of the order of $10^{-4}$\,G according to the estimate for the equipartition magnetic field of the radio component A, located at a projected distance of $\sim$350\,pc from the core of NGC\,1068 \citep{1983MNRAS.202..647P}. No significant short-term variability is expected from such a large emitting zone.

The stationary emission of a blob of plasma is modelled here through leptonic processes. The macrophysics of the blob is described by its bulk Doppler factor $\delta_\mathrm{b}$, radius $r_\mathrm{b}$, and strength of the tangled magnetic field $B$ filling it up. The energy distribution of the leptons is described by a broken power-law with the density normalisation $K$, the indices $n_1$ and $n_2$, and the Lorentz factors of the individual particles $\gamma_\mathrm{min}$, $\gamma_\mathrm{break}$ and $\gamma_\mathrm{max}$, as follows:

\begin{equation}
  N_\mathrm{e}(\gamma)=
  \begin{cases}
    K \gamma^{-n_1} & \text{if } \gamma_\mathrm{min} \leqslant \gamma \leqslant \gamma_\mathrm{break} \\
    K \gamma_\mathrm{break}^{n_2-n_1} \gamma^{-n_2} & \text{if } \gamma_\mathrm{break} \leqslant \gamma \leqslant \gamma_\mathrm{max}
  \end{cases}
  \quad [\mathrm{cm}^{-3}]
  \label{eq-EED}
\end{equation}

These leptons emit through synchrotron self-Compton \citep[SSC, see e.g.][]{1965ARA+A...3..297G} and EIC processes. The latter is described by the fraction of the core luminosity $\tau L_\mathrm{nuc}$ reprocessed by the emitting source \citep[see][]{1996ApJ...463..555I}, the temperature $T$ of the seed radiation field described as a blackbody, and the distance $R$ between the external radiation field and the re-emitting blob. Second order SSC emission \citep{1967MNRAS.137..429R} is also accounted for in this model. More details on the model can be found in \citet{2001A+A...367..809K}, \citet{2008A+A...478..111L}, and \citet{2009PhDT.........JPL}. We use $\gamma_\mathrm{min}=1$ in the following.

In Fig.~\ref{fig-model-NGC1068}, we present such a model with synchrotron emission responsible for the radio emission, while the \textit{Fermi}/LAT data are interpreted as EIC emission with the infrared emission providing the seed photons, from a multi-temperature blackbody. We also show the contribution from the SSC process, which is negligible compared to the EIC emission at the highest energies. We note that the contribution of second order SSC is negligible in the case of our interpretation of the SED of NGC\,1068. The model parameters are summarised in Table~\ref{tab-model_parameters}.

The hard X-ray spectrum observed with \textit{INTEGRAL} does not seem to fit in this picture and we propose that this emission originates from EIC processes on another population of leptons, within hot plasma located in the vicinity of the accretion disc. The seed photons would then originate from the accretion disc itself as is usually invoked to explain the X-ray emission of Seyfert galaxies. Evidence for an accretion disc comes from the fact that the soft X-ray spectrum of NGC\,1068, e.g. as observed with {\it XMM}-Newton, is dominated by thermal reflection emission \citep[see e.g.][]{2002ApJ...575..732K,2006MNRAS.368..707P}. We checked that the synchrotron and the SSC emission from this component is negligible compared to the one from the large outflow described above.

It should be noted that imaging atmospheric \v{C}erenkov telescopes have the ability to strongly constrain the highest energy part of the particle energy distribution. If the maximum energy of the leptons is $\gamma_\mathrm{max} \gtrsim 5 \times 10^6$, a significant signal could be detected from NGC\,1068 with H.E.S.S., VERITAS, MAGIC, or the future \v{C}erenkov Telescope Array (CTA) \citep{2010arXiv1008.3703C}.

An alternative solution from our model would consist in interpreting the hard X-ray emission from \textit{INTEGRAL} and the high energy \g-ray emission from \textit{Fermi}/LAT as originating from the same spectral component. In this case, the overall high energy emission would be due to EIC process from the same lepton population. The spectral shapes of both \textit{INTEGRAL} and \textit{Fermi} data then strongly constrain the parameters on the lepton energy distribution, for which we obtain $n_1=2.0$, $n_2=3.6$, $\gamma_\mathrm{break}=700$ and $K=450$\,cm$^{-3}$. However, such a particle energy distribution can not account correctly for the SED observed in the radio domain.


In Fig.~\ref{fig-SED-NGC4945}, we show the SED of NGC\,4945, including the \textit{Fermi}/LAT and the \textit{INTEGRAL} spectra we derived in Section~\ref{sec-fermi_analysis} and \ref{sec-integral_analysis}, respectively. For comparison, we also show the data of the starburst galaxy NGC\,253, as taken from the NED, along with the \textit{Fermi}/LAT spectrum as published in \citet{2010ApJ...709L.152A}. Clearly, the two objects have very similar broadband SEDs, which match almost perfectly, strengthening the idea that the high energy \g-ray emission from NGC\,4945 could be well explained only in terms of starburst activity. NGC\,4945 should then be detectable in the very high energy domain, and especially with H.E.S.S. given its coordinates, at a flux level similar to NGC\,253. However, one difference between NGC\,4945 and NGC\,253 resides in their hard X-ray emission. While NGC\,253 is not thought to harbour a central AGN, NGC\,4945 exhibits a characteristic Seyfert emission, which is modelled as above for NGC\,1068. The corresponding parameters are also included in Table~\ref{tab-model_parameters}.

%

\section{Conclusion}

We reported the detection of the Seyfert 2 galaxy NGC\,1068 in the high energy (100\,MeV--100\,GeV) domain with the {\it Fermi}/LAT, and compared the analysis of this source with the one of NGC\,4945 obtained by extending the observing period to 1.6 years. The object 1FGL\,J0242.7$+$0007 was reported as a \g-ray source without association in the 1FGL catalogue. We can now associate it quite firmly to NGC\,1068 with an overall detection significance of 8.4\,$\sigma$. The data do not show any significant variability over the whole period.

The high energy \g-ray spectrum of NGC\,1068 is consistent with a power-law of index $\Gamma = 2.31 \pm 0.13$ and a flux density of \fluxone. Compared to the \g-ray luminosities of M\,82 and NGC\,253, whose high-energy emission is dominated by starburst activity, we find a too high \g-ray luminosity of NGC\,1068 to be explained only by starburst activity. We thus propose a leptonic scenario to interpret the high energy emission in terms of external inverse Compton process from an outflowing relativistic wind launched by the central AGN.

NGC\,4945 is detected at a 9.2\,$\sigma$ level in \textit{Fermi}/LAT data, and its \g-ray spectrum is best described by a power-law of index $\Gamma = 2.31 \pm 0.10$ and a flux density of \fluxtwo. This object has a very similar multi-wavelength SED compared to NGC\,253, and based on its radio, infrared and \g-ray luminosities, its high energy \g-ray emission is most likely due to starburst activity.

If high energy \g-ray emission due to AGN activity is confirmed in other Seyfert 2 galaxies, this would mark the discovery of yet a new class of high-energy \g-ray emitters. New data from \textit{Fermi}/LAT in the coming years will be extremely valuable in this regard.

%

\begin{acknowledgements}
  We thank our colleague Dr.~Christian Farnier for valuable discussions. This research has made use of NASA's Astrophysics Data System (ADS), of the SIMBAD database, operated at CDS, Strasbourg, France, and of the NASA/IPAC Extragalactic Database (NED) which is operated by the Jet Propulsion Laboratory, California Institute of Technology, under contract with the National Aeronautics and Space Administration.
  
  J.-P.~L. would like to dedicate this work to the memory of his missed friend Jean-Claude Rouffignat.
\end{acknowledgements}

\bibliographystyle{aa}  
\bibliography{Fermi_NGC1068_NGC4945.bbl}


\end{document}